Commentary

# CONSERVATION OF NUCLEAR SSR LOCI REVEALS HIGH AFFINITY OF *QUERCUS INFECTORIA* SSP. *VENERIS* A. KERN (FAGACEAE) TO SECTION *ROBUR*

Ch. Neophytou[1,*], A. Dounavi[1] and F.A. Aravanopoulos[2]

1) Forest Research Institute – Baden-Württemberg, Wonnhaldestr. 4, D-79100, Freiburg
2) Faculty for Forestry and Natural Environment, Aristotle University of Thessaloniki, P.O. BOX. 238, Thessaloniki, Greece

*****Author for correspondence:

e-mail: charalambos.neophytou@forst.bwl.de
Telephone Number: +49 761 4018161
Fax Number: +49 761 4018133

**Abstract**

Conservation of 16 nuclear microsatellite loci, originally developed for *Quercus macrocarpa* (section *Albae*), *Q. petraea*, *Q. robur* (section *Robur*) and *Q. myrsinifolia*, (subgenus *Cyclobalanopsis*) was tested in a *Q. infectoria* ssp. *veneris* population from Cyprus. All loci could be amplified successfully and displayed allele size and diversity patterns that match those of oak species belonging to the section *Robur*. At least in one case, limited amplification and high levels of homozygosity support the occurrence of “null alleles”, caused by a possible mutation in the highly conserved primer areas, thus hindering PCR. The sampled population exhibited high levels of diversity despite the very limited distribution of this species in Cyprus and extended population fragmentation. Allele sizes of *Q. infectoria* at locus QpZAG9 partially match those of *Q. alnifolia* and *Q. coccifera* from neighboring populations. However, sequencing showed homoplasy, excluding a case of interspecific introgression with the latter, phylogenetically remote species. *Q. infectoria* ssp. *veneris* sequences at this locus were concordant to those of other species of section *Robur*, while sequences of *Quercus alnifolia* and *Quercus coccifera* were almost identical to *Q. cerris*.

Abbrevations: SSR, Single Sequence Repeat; PCR, Polymerase Chain Reaction; A, adenine; G, guanine; C, cytocine; T, thymine.

## Introduction

The genus *Quercus* is one of the most ecologically and economically important taxa covering large parts of the northern hemisphere. Oak species distribution varies from Mediterranean sclerophyllous ecosystems, to temperate deciduous forests. Oak reproductive patterns and their tendency to intercross made them one of the most favorite subjects of evolutionary studies since Darwinian times.

Microsatellites or Simple Sequence Repeats (SSRs) have been widely used for population genetic studies during the last decade, including oak species (Kampfer et al., 1998). Neutrality, codominance and high levels of polymorphism are properties that make them an applicable tool for dealing with numerous research problems, such as genetic diversity, paternity analyses and spatial genetics studies.

*Quercus infectoria* ssp. *veneris* (gall oak) is a semi-deciduous tree species distributed in an area ranging from the islands of the Aegean Sea to SW. Iran (Zohary, 1973; Meikle, 1985), with a Mediterranean to semi-arid climate. It is considered as a multipurpose tree and its uses include production of timber, dyes, acorns and medical extracts. Natural populations are threatened from limitation and fragmentation of its habitats due to human activities (Nimri et al., 1999; Christou, 2000). In Cyprus, gall oak is nowadays restricted to small fragmented populations. These are believed to be remnants of extended woodlands formerly covering the southern part of the Troodos Mountains (Meikle 1977; Christou 2000). From the ecological point of view, gall oak plays an important role by forming a special phytosociological association characterized by a rich flora and developing deep forest soils (Barbéro and Quézel 1979). Gall oak ecosystems are included in the European Ecological Network Natura 2000.

The taxonomical and phylogenetic status of gall oak still remains unclear. A close affinity or synonymy with *Quercus pubescens* (section *Robur* according to Krüssmann (1998)) is partially claimed (Schwarz, 1993; Schirone and Spada 2000). However, inferences about the phylogeny of *Q. infectoria* are based on morphological traits (Schwarz, 1936) and research with molecular markers is still lacking.

In this paper we tested the conservation and applicability of a series of nuclear SSR markers developed for *Q. petraea*, *Q. robur* and *Q. myrsinifolia.* In locus ssrQpZAG9, homoplasy has already been described between oak species belonging to different sections (Curtu et al., 2004). We tested possible homoplasy by sequencing alleles of the same length among the above species using additionally samples of *Q. alnifolia* and *Q. coccifera*, which are the other two oak species growing in Cyprus and belong to section *Cerris* (sensu Manos et al. 1999).

Knowledge about the transferability of SSR markers to gall oak is the basis for using them in further genetic studies. Pertinent research could serve in the conservation of endangered gall oak populations, such as those of Cyprus. In addition, in light of global warming, it is important to enhance the knowledge about an oak species potentially closely related to sessile (*Q. petraea*) and pedunculate (*Q. robur*) oak, which play an important role in European forestry. The present study provides some first indications about the taxonomical relatedness of gall oak with the aforementioned species.

**Materials and Methods**

Leaf material was collected from 50 mature trees from in the southern part of Troodos Mountains in Cyprus. Sampled trees were randomly collected from six fragmented stands. Trees were located at least 30 m apart from each other, in order to avoid family structures which may cause bias during calculating diversity measures. Depending on the size of the stand, sample sizes varied between 3 and 13 each.

DNA was extracted according Dumolin *et al*. (1995) with some modifications and an additional overnight treatment with RNAseA (Qiagen) at 37°C. Transferability of 16 nuclear polymorphic microsatellite loci was tested. Loci MSQ3, MSQ4 and MSQ13 were initially developed in *Q. macrocarpa* (Dow et al., 1995); locus QM50-3M in *Q. myrsinifolia* (Isagi and Suhadono, 1997); loci ssrQpZAG9, ssrQpZAG15, ssrQpZAG46, ssrQpZAG104 and ssrQpZAG110 were initially described in *Q. petraea* (Steinkellner et al., 1997a); loci ssrQrZAG7, ssrQrZAG30, ssrQrZAG65, ssrQrZAG87, ssrQrZAG96, ssrQrZAG101 and ssrQrZAG102 were originally characterized in *Q. robur* (Kampfer et al., 1998). Tested loci are scattered among 10 out of 12 linkage groups described for *Q. robur* and *Q. petraea* (Barreneche et al., 1998; Scotti-Saintagne et al., 2004), corresponding to respective numbers of genomic chromosomes.

PCRs were performed either using a PTC-200 Gradient Cycler (MJ Research, Inc.) or an ABI-2720 Thermal Cycler (Applied Biosystems). The volume of the reaction mixture was 25µl containing 10ng of template DNA and 1x Reaction Buffer (Invitrogen, USA).  The latter contained 20mM tris-HCl (pH 8,0), 40 mM NaCl, 2mM

Sodium Phosphate, 0,1 mM EDTA, 1mM DTT, stabilizers and 50% (v/v) glycerol); 1,5 mM $MgCl_2$, 200 mM of each dNTP, 1 unit Platinum Taq DNA Polymerase (Invitrogen, USA) and 0,2 mM Primer (Biomers, Germany). The PCR programs were based on Kampfer et al. (1998) with the addition of 15 s elongation steps in each cycle. PCR products were scored in an automated sequencer (ABI PRISM 3100, Applied Biosystems) and alleles were detected with GeneMapper v4.0 Software (Applied Biosystems). Data were then processed with the Popgene 32 (Yeh et al., 1997) software.

Separation of SSR alleles originating from locus ssrQrZAG9 for sequencing was carried out on a 2% agarose gel. Fragments were subsequently purified using a QiaQuick Gel-Extraction Kit (Qiagen). For the sequencing reaction the 3.1 BigDye Terminator v3.1 Sequencing Standard Kit (Applied Biosystems) was used. Eight µl Terminator Ready Reaction Mix, 2-4µl (10 ng) of template DNA and 1µl (5µM) primer were mixed in 20µl of reaction solution. The sequencing reaction took place in an ABI-2720 thermal cycler (Applied Biosystems). The program included 1 stage of 96°C for 3 min, 30 cycles of 96°C for 10 s, 50°C for 10 s and 60°C for 4 min. Removal of residual dye terminators was carried out using Centri-Sep spin columns (Genaxxon). Sequencing electrophoresis run on an ABI PRISM 3100 Genetic Analyzer (Applied Biosystems) and results were analyzed with the Sequencing Analysis 5.1.1 software (Applied Biosystems). Further data processing was carried out using the SepScape 2.5 (Applied Biosystems) software.

## Results and Discussion

### *Conservation of SSR loci*

All tested microsatellite loci could be successfully amplified in gall oak and showed diploid patterns. In general, they exhibited high levels of polymorphism and fragment sizes matched those initially measured in the source species (Table 1). In many loci, allelic richness exceeded this found in the source species. For example, for markers MSQ4 and MSQ13, 19 and 18 alleles were respectively found in 49 tested trees, while Dow et al. (1995) found 11 and 12 alleles after testing 61 bur oaks. A similar result was found in locus QM50-3M where 5 alleles were detected in 20 *Q. myrsinifolia* individuals (Isagi and Suhadono, 1997), while in this study 22 alleles were detected in 46 gall oaks. Effective number of alleles is also relatively high (Table 1), showing that different variants at each locus are represented in high frequencies within the sample. Additionally, the majority of the loci displayed high levels of observed heterozygosity and low fixation indexes (Table 1). The aforementioned results argue for a high genetic variability of *Q. infectoria*. Taking into account the loci tested belong to 10 out of 12 genomic chromosomes, it could be postulated that variability reflected in our data might be genome wide. Population genetics measures also provide some first indication, that this species kept high levels of genetic diversity, despite long time fragmentation and limitation of its habitats in Cyprus.

Nevertheless, loci MSQ3, QpZAG15, QrZAG65 and QrZAG102 displayed markedly large difference between observed and expected heterozygosity and consequently high fixation indexes. At locus MSQ3, initially developed for *Quercus macrocarpa*, poor or no amplification products were identified in a number of individuals. Only five alleles were scored, while allele size range was apparently narrower than in *Q. macrocarpa*. In 35 gall oak individuals, only one allele was detected, while in the rest 15 samples PCR was unsuccessful. Similar patterns were observed for markers QpZAG15, QrZAG65 and QrZAG102, however with a much higher number of detected alleles and lower number of unsuccessful PCRs. Mutations in primer binding sites result in the occurrence of “null alleles” (Vornam et al., 2004), a possible explanation for this study’s results. It can be suggested that these four loci should be avoided in genetic diversity studies of *Q. infectoria* populations.

*Sequencing analysis at microsatellite locus QpZAG9*

At locus QpZAG9 most allele sizes varied from 176 bp to 206 bp showing a normal distribution, which is typical for oak species belonging to section *Robur* (Steinkellner et al., 1997a). However, rare alleles with sizes of 248, 254, 255 and 258 bp emerged in heterozygous individuals in our study (Figure 1). *Q. cerris*, as other species classified to section *Cerris*, for instance *Q. ilex*, *Q. suber*, *Q. alnifolia* and *Q. coccifera* (Steinkellner et al., 1997b; Hornero et al., 2001; Soto et al., 2003; Neophytou, unpublished data), exclusively possess alleles with sizes ranging from 223 to 264 bp. Curtu et al. (2004) also detected alleles of this size range in species classified in section *Robur* in a mixed oak stand in Romania. Alleles were present in very low frequencies and always in heterozygous condition. Nevertheless, homoplasy was revealed, since different sequences were detected between equally sized alleles occurring in species that belong to different sections. Sequencing of allele “248”, using probes of *Q. alnifolia* and *Q. coccifera* from neighboring stands was performed, in order to investigate if the presence of rare alleles in this study is a result of introgressive hybridization or another case of homoplasy. It should be mentioned that allele size “248” was the only one observed in all species, among more than a 100 individuals of each *Q. alnifolia* and *Q. coccifera* (Neophytou, unpublished data).

Results of sequencing confirm that, despite the identical size, allele “248” in *Q. infectoria* possesses considerably different sequence compared with *Q. alnifolia* and *Q. coccifera*, while it differs from allele “247” in *Q. robur* from Curtu et al. 2004 by only one point mutation. *Q. alnifolia* and *Q. coccifera* share almost identical sequences for allele “248”, being different in one a single point mutation.

Differences between *Q. infectoria* and the group of *Q. alnifolia* and *Q. coccifera* include indels, point mutations and repeat number at the microsatellite region. In particular, different number of repeats is observed in the microsatellite area where *Q. alnifolia* and *Q. coccifera* possess 10, and *Q. infectoria* possesses 7 repeats (Figure 2). Single nucleotide polymorphisms differentiating the former species from the latter occur at positions 89 (C versus T), 173 (G versus C) and 217 (A versus C), with the sequence of *Q. infectoria* matching this of *Q. robur* reported by Curtu et al. (2004). A point mutation at position 120 is the only difference between *Q. infectoria* (C) and *Q. robur* (G) in the SSR flanking region. A transversion can be observed at position 261, where *Q. coccifera* possesses a thymine, while all other species possess an adenine. Moreover, at position 98 a cytosine was found in all studied species, although all *Q. cerris* samples of Curtu et al. (2004) are characterized by a thymine. Other differences observed include insertions at positions 114 and 141 for *Q. infectoria*, which shows the same pattern as *Q. robur*. *Q. alnifolia* and *Q. coccifera* exhibit at this region the same pattern as *Q. cerris* in the study of Curtu et al. (2004). *Q. infectoria* shows large insertions, with sequences almost identical to *Q. robur*. A short insertion distinguishing *Q. infectoria* from *Q. alnifolia* and *Q. coccifera* occurs at position and 86. Here, the former species follows the pattern of section *Robur*, while *Q. alnifolia* and *Q. coccifera* deviate from *Q. cerris*, which possesses a deleterious variant.

Sequencing results in general show striking similarities between *Q. infectoria* and other species of the section *Robur* in Europe. Amplification patterns of the tested microsatellite loci also argue for a classification of the gall oak to the same taxonomical unit. On the other hand, sequences of *Q. alnifolia* and *Q. coccifera* do not absolutely match the sequence of *Q. cerris* as reported in Curtu et al. (2004). This may suggest a longer time of divergence, between Mediterranean evergreen oaks and other "*Cerris*" oaks and may justify the taxonomical distinction supported by some authors (Schwarz, 1936; Camus, 1938). The evolutionary history of *Q. infectoria* might be similar to oaks of section *Robur*, which migrated from an evolutionary centre in America towards Europe during the Tertiary. In contrast, evergreen oaks belonging to the section *Cerris* occupied the Mediterranean basin after emerging from an evolutionary centre in SW China during the late Tertiary (Axelrod, 1983; Zhou, 1992; Manos et al., 1999). Absence of gene flow between *Q. infectoria* on one hand and *Q. coccifera* and *Q. alnifolia* on the other, as shown in our study, is in concordance with that phylogenetic classification, as strong reproductive barriers and gametic incompatibility exist between the two groups after a long period of geographic isolation (Boavida et al., 2001).

## Conclusions

The description of a set of highly variable nuclear microsatellite markers amplified in *Q. infectoria* ssp. *veneris* in this study provides an efficient tool for studying genetic diversity and mating systems in this species. This is of particular interest in the context of conservation of this species in Cyprus, where its populations only consist of small fragmented stands or isolated trees. Additionally, our data support a close relationship of this species with other European oak species of section *Robur*, elucidating the phylogenetic status of *Q. infectoria* ssp. *veneris*. The high genetic affinity of *Q. infectoria* ssp. *veneris* to oaks of section *Robur* could suggest the inclusion of *Q. infectoria* ssp. *veneris* in this section. Populations of this species are adapted to climatic conditions much warmer and drier than in those of the main distribution area of section *Robur* in Europe. Therefore gall oak is an interesting candidate for future exploitation in European forestry in light of global warming and climatic change.

## Acknowledgments

We are grateful to Constantinos Kounnamas and Costas Kadis for their advice and assistance during planning and carrying out the collections. This research was conducted in partial fulfilment for the degree of the Albert-Ludwigs University of Freiburg regarding the senior author.

**References**

Axelrod DI (1983) Biogeography of oaks in the Arcto-Tertiary province. Ann Missouri Bot Gard 70: 629-657

Barbéro M, Quézel P (1979) Contribution à l' étude des groupements forestiers de Chypre. Doc phytosociologiques IV: 9-34

Barreneche T, Bodenes C et al. (1998) A genetic linkage map of *Quercus robur* L. (pedunculate oak) based on RAPD, SCAR, microsatellite, minisatellite, isozyme and 5S rDNA markers. Theor Appl Gen 97: 1090-1103

Boavida LC, Silva JP et al. (2001) Sexual reproduction in the cork oak (*Quercus suber* L). - II. Crossing intra- and interspecific barriers. Sex Pl Reprod 14: 143-152

Camus A (1938). Les chênes. Lechevallier, Paris

Christou A (2000). Cyprus, Country Report. Presented in EUFORGEN Mediterranean Oak Network, Antalya 12-14 October 2000.

Curtu AL, Finkeldey R, et al. (2004). Comparative sequencing of a microsatellite locus reveals size homoplasy within and between European oak species (*Quercus* spp.). Pl Mol Biol Reporter 22: 339-346

Dow BD, Ashley MV, Howe HR (1995). Characterization of highly variable (GA/CT)n microsatellites in the bur oak, *Quercus macrocarpa*. Theor Appl Gen 91: 137-141

Dumolin S, Demesure B, Petit RJ (1995) Inheritance of chloroplast and mitochondrial genomes in pedunculate oak investigated with an efficient PCR method. Theor Appl Gen 91: 1253-1276

Hornero J, Gallego FJ, et al. (2001). Testing the conservation of *Quercus* spp. microsatellites in the cork oak, *Q. suber* L. Silvae Genetica 50: 162-167

Isagi Y, Suhandono S (1997). PCR primers amplifying microsatellite loci of *Quercus myrsinifolia* Blume and their conservation between oak species. Mol Ecol 6: 897-899

Kampfer S, Lexer C, et al. (1998). Characterization of (GA)n microsatellite loci from *Quercus robur*. Hereditas 129: 183-186

Krüssmann G (1978) Handbuch der Laubgehölze, Paul Parey Verlag, Berlin

Manos PS, Doyle JJ, et al. (1999). Phylogeny, biogeography, and processes of molecular differentiation in *Quercus* subgenus *Quercus* (Fagaceae). Mol Phylog Evol 12: 333-349

Meikle RD (1977). Flora of Cyprus. Bentham Moxon Trust, Royal Botanic Gardens, London

Meikle RD (1985). Flora of Cyprus. Bentham Moxon Trust, Royal Botanic Gardens, London

Nimri, LF, Meqdam MM, et al. (1999). Antibacterial activity of Jordanian medicinal plants. Pharmaceutical Biology 37: 196-201

Schirone B, Spada F (2000). Some remarks on the conservation of genetic resources of Meditderranean oaks. Presented in EUFORGEN - Mediterranean Oaks Network, Antalya 12-14 October 2000

Schwarz, O (1936) Entwurf einem naturlichen system der Culpuliferen und der Gattung *Quercus* L. Notizbl Bot Gart Berlin 13: 1-22

Schwarz, O (1993) *Quercus* L. In: Flora europea, vol. I, 2nd edn. Tutin TG, Burger VH, Valentine DH, Walters SM, Webb DA eds.

Scotti-Saintagne C, Mariette S et al. (2004). Genome scanning for interspecific differentiation between two closely related oak species [*Quercus robur* L. and *Q petraea* (Matt.) Liebl.]. Genetics 168: 1615-162

Soto, A, Lorenz Z et al. (2003) Nuclear microsatellite markers for the identification of *Quercus ilex* L. and *Q. suber* L. hybrids. Silvae Genetica 52: 63-66

Steinkellner H, Lexer C et al. (1997) Conservation of (GA)(n) microsatellite loci between *Quercus* species. Mol Ecol 6: 1189-1194

Steinkellner H., Fluch S. et al. (1997) Identification and characterization of (GA/CT)n- microsatellite loci from *Quercus petraea*. Pl Mol Biol 33: 1093-1096

Vornam B, Decarli N, et al. (2004) Spatial distribution of genetic variation in a natural beech stand (*Fagus sylvatica* L.) based on microsatellite markers. Conserv Gen 5: 561-570

Yeh, F. Boyle, TJB et al. (1997) Population genetic analysis of co-dominant and dominant markers and quantitative traits. Belgian Jour Bot 129(157)

Zhou, ZK (1992) Origin, phylogeny and dispersal of *Quercus* from China. Acta Bot Yunnanica 14: 227-236

Zohary, M (1973) Geobotanical Foundations of the Middle East. Gustav Fischer Verlag, Stuttgart

Tables:

| Locus | LG | Source species | | *Quercus infectoria* | | | | | |
|---|---|---|---|---|---|---|---|---|---|
| | | Al. size | n | Al. size | n | $n_e$ | $H_o$ | $H_e$ | F |
| | | | | | | | | | |
| MSQ3 | | 191-231 | 20 (61) | 199-205 | 5 (34) | 3,56 | 0,088 | 0,730 | 0,879 |
| MSQ4 | 4 | 203-227 | 11 (61) | 200-233 | 19 (49) | 8,98 | 0,857 | 0,898 | 0,046 |
| MSQ13 | 6 | 222-246 | 12 (61) | 195-246 | 18 (49) | 9,55 | 0,857 | 0,905 | 0,053 |
| QM50-3M | | 253 | 4 (20) | 250-293 | 22 (46) | 10,32 | 0,783 | 0,913 | 0,142 |
| ssrQrZAG7 | 2 | 115-153 | 10 (6) | 113-142 | 15 (49) | 5,79 | 0,776 | 0,836 | 0,072 |
| ssrQpZAG9 | 7 | 182-210 | 11 (28) | 176-258 | 17 (49) | 9,47 | 0,837 | 0,904 | 0,074 |
| ssrQpZAG15 | 9 | 108-152 | 11 (45) | 101-132 | 20 (40) | 10,67 | 0,500 | 0,918 | 0,455 |
| ssrQrZAG30 | 12 | 172-248 | 10 (5) | 172-248 | 19 (48) | 8,13 | 0,771 | 0,886 | 0,130 |
| ssrQpZAG46 | 2 | 190-222 | 8 (24) | 188-219 | 19 (40) | 11,43 | 0,725 | 0,924 | 0,215 |
| ssrQrZAG65 | 10 | 261-302 | 9 (6) | 254-324 | 24 (40) | 13,79 | 0,325 | 0,934 | 0,652 |
| ssrQrZAG87 | 2 | 110-181 | 8 (6) | 95-118 | 9 (49) | 3,88 | 0,776 | 0,750 | -0,035 |
| ssrQrZAG96 | 10 | 135-194 | 8 (6) | 137-176 | 19 (49) | 12,25 | 0,816 | 0,928 | 0,121 |
| ssrQrZAG101 | 1 | 136-160 | 7 (6) | 131-171 | 17 (49) | 10,67 | 0,857 | 0,916 | 0,064 |
| ssrQrZAG102 | 3 | 195-276 | 6 (6) | 194-280 | 19 (41) | 5,49 | 0,463 | 0,828 | 0,441 |
| ssrQpZAG104 | 2 | 176-196 | 9 (25) | 175-217 | 17 (49) | 7,26 | 0,755 | 0,871 | 0,133 |
| ssrQpZAG110 | 8 | 206-262 | 7 (40) | 198-238 | 18 (46) | 10,61 | 0,870 | 0,916 | 0,050 |

Table 1 – Comparison of allele richness between source species and *Q. infectoria*. Linkage group is shown for each locus. Allele sizes and number of alleles (n; in parentheses number of tested individuals) are presented for source species and *Q. infectoria* from our study. Effective number of alleles ($n_e$), observed ($H_o$) and expected heterozygosity ($H_e$) and fixation index ($F=1-H_o/H_e$) for *Q. infectoria* are additionally indicated. Data for source species were taken from studies cited in Materials and Methods.

Captions of Figures

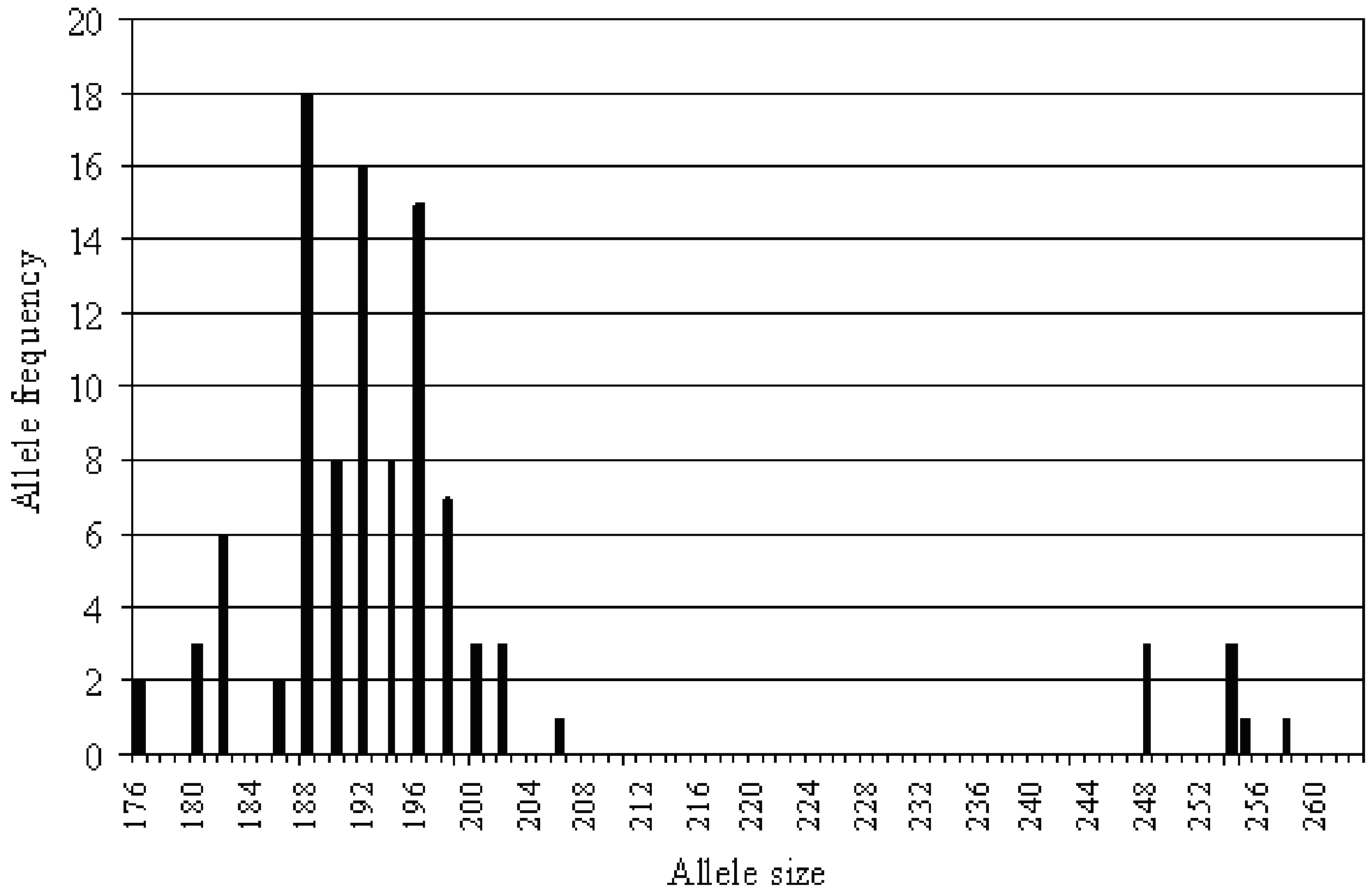

Figure 1 – Allele frequencies for locus QpZAG9 in *Q. infectoria* from our study. Most alleles emerge within the size range 176 to 206, while scarce alleles with sizes between 248 and 258 occur in heterozygous condition.

```
10         20        30        40        50        60        70        80        90        100       110       120       130       140
          ....|....|....|....|....|....|....|....|....|....|....|....|....|....|....|....|....|....|....|....|....|....|....|....|....|....|....|....|....|

Q.inf.    GCAATTACAGGCTAGGCTGG-----GTGGTGATTGGTGATTGGATTTGAGAGAGAGAGAGAG--------------------------TCAGTCAGTCCCTTCTCTGGGCTGGGGTTGGCTTGG-----CTGGCCTAATAATAAT
Q.aln.    GCAATTACAGGCTAGGCTGG-----GTGGTGATTGGTGATTGGATTTGAGAGAGAGAGAGAGAGAGA------------------CAGCCAGTCAGTCCCTTCTCTGGGCTGG--------------------CCTAATAAT---
Q.coc.    GCAATTACAGGCTAGGCTGG-----GTGGTGATTGGTGATTGGATTTGAGAGAGAGAGAGAGAGAGA------------------CAGCCAGTCAGTCCCTTCTCTGGGCTGG--------------------CCTAATAAT---

Q.rob.    GCAATTACAGGCTAGGCTGG-----GTGGTGATTGGTGATTGGATTTGAGAGAGAGAGAG----------------------------TCAGTCAGTCCCTTCTCTGGGCTGGGCTGGGGTTGG-----CTGGCCTAATAATAAT
Q.cer.    GCAATTACAGGCTAGGCTGGGCTGGGTGGTGATTGGTGATTGGATTTGAGAGAGAGAGAGAGAGAGAGAGAGGAGAGAGAGAGAG---TCAGTCAGTTCCTTCTCTGGGCTGG--------------------CCTAATAAT---

             150       160       170       180       190       200       210       220       230       240       250       260       270       280
          ....|....|....|....|....|....|....|....|....|....|....|....|....|....|....|....|....|....|....|....|....|....|....|....|....|....|....|.

Q.inf.    TGTATTTGGTTTAGAGGGGAAGGAAAGCAAG-CAGCAAATCGAAAATGAGCGTGGTTTCGGGTGTGATATCCCGCCAAGTATTGCCGGC-ATGCGGTAGTCTCTGTT-TCTTTTGCCCCGCCATGAGGGCTAGGTCCAGAC
Q.aln.    TGTATTTGGTTTAGAGGGGAAGGAAAGGAAG-CAGCAAATCGAAAATGAGCGTGGTTTCGGGTGTGATATCACGCCAAGTATTGCCGGC-ATGCGGTACTCTCTGTT-TCTTTTGCCCCGCCATGAGGGCTAGGTCCAGAC
Q.coc.    TGTATTTGGTTTAGAGGGGAAGGAAAGGAAG-CAGCAAATCGAAAATGAGCGTGGTTTCGGGTGTGATATCACGCCAAGTATTGCCGGC-ATGCGGTACTCTCTGTT-TCTTTTGTCCCCCCATGAGGGCTAGGTCCAGAC

Q.rob.    TGTATTTGGTTTAGAGGGGAAGGAAAGCAAG-CAGCAAATCGAAAATGAGCGTGGTTTCGGGTGTGATATCCCGCCAAGTATTGCCGGC-ATGCGGTAGTCTCTGTT-TCTTTTGCCCCGCCATGAGGGCTAGGTCCAGAC
Q.cer.    TGTATTTGGTTTAGAGGGGAAGGAAAG--------CAAATCGAAAATGAGCGTGGTTTCGGGTGTGATATCACGCCAAGTATTGCCGGC-ATGCGGTAGTCTCTGTT-TCTTTTGCCCCGCCATGAGGGCTAGGTCCAGAC
```

Figure 2 – Sequence alignments of allele “248” at locus ssrQpZAG9 in *Q. infectoria, Q. alnifolia* and *Q. coccifera* from our study and of allele “247” in *Q. robur* and “251” in *Q. cerris* from Curtu et al. 2004.